\documentstyle[12pt]{article}
\newcommand{\be}{\begin{equation}}
\newcommand{\bea}{\begin{eqnarray}}
\newcommand{\eea}{\end{eqnarray}}
\newcommand{\ba}{\begin{array}}
\newcommand{\ea}{\end{array}}
\newcommand{\ee}{\end{equation}}
\newcommand{\bal}{\mbox{\boldmath$\alpha$}}
\newcommand{\bde}{\mbox{\boldmath$\delta$}}

\newcommand{\qq}{\mbox{\boldmath$q$}}
\newcommand{\m}{\mbox{\boldmath$g$}}
\expandafter\ifx\csname mathbbm\endcsname\relax

\newcommand{\bone}{\mbox{1\hspace{-0.65em}1}}
\else

\newcommand{\bone}{{\mathbbm{1}}}
\fi
\begin{document}
\begin{titlepage}
\begin{flushleft}
       \hfill                      IPM-95-117\\
\end{flushleft}
\vspace*{3mm}
\begin{center}
{\LARGE The Moduli Space of the $N=2$ Supersymmetric $G_{2}$ Yang-Mills 
Theory\\}

\vspace*{12mm}
{\large M. Alishahiha, \,\, F. Ardalan } \\

{\it Institute for Studies in Theoretical Physics and Mathematics, \\
 P.O.Box 19395-1795, Tehran, Iran } \\
{\it  Department of Physics, Sharif University of Technology, \\
\it  P.O.Box 11365-9161, Tehran, Iran \/}\\
\vspace*{5mm}
{\large F. Mansouri } \\
{\it Department of Physics, University of Cincinnati, \\
Cincinnati, Ohio, 45221, U. S. A \/}
\end{center}
\vspace*{15mm}

\begin{abstract}

We present the hyper-elliptic curve describing the moduli space of the N=2
supersymmetric Yang-Mills theory with the $G_2$ gauge group. The exact 
monodromies and the dyon spectrum of the theory are determined.
It is verified that the recently proposed solitonic equation is also 
satisfied by our solution.

\end{abstract}

\end{titlepage}

In an important developement in four dimensional quantum field theory, 
recently
a number of exact results have been obtained \cite{ALV}. In particular, for
$N=2$ supersymmetric gauge field theory, the exact low energy prepotential has
been determined with the help of an auxiliary hyper-elliptic curve, which 
among
other things allows the determination of the massless dyon spectrum of these
theories.

This program, initially originated by Seiberg and Witten for the gauge group
$SU(2)$ \cite{SW}, has been carried out for $SU(N)$ \cite{LA}, $SO(2n+1)$ 
\cite{DA}
and $SO(2n)$ \cite{BR} so far. In this letter we will describe the solution
for the group $G_2$.

The low energy effective action for the $N=2$ gauge theory, written in terms
of the $N=1$ fields is:
\be
{ 1 \over 4 \pi} Im \left( \int d^{4} \theta { \partial {\cal F} (A)
\over \partial A^{i} } \bar{A}^{i} + \int d^{2} \theta {1 \over 2}
{ \partial ^{2} {\cal F} ( A ) \over \partial A^{i} \partial A^{j}} 
W^{i}_{\alpha} W^{j}_{\alpha} \right)   ,   \label{verkan}
\ee
where $A^{i}$ are the N=1 chiral field multiplets and $W^{i}_{\alpha}$ 
are the vector multiplets all in the adjoint representation. 
The prepotential ${\cal F}$, which at the classical level is
\be\ba{ll}
{\cal F}={1 \over 2} \tau_{cl} A^{2} ,& \tau_{cl}={\theta \over 2\pi}+
i{4\pi \over g^2},
\ea\ee
gets the one-loop correction
\be
{\cal F}={i \over 2\pi} A^2 ln({A^2 \over \Lambda^2}).
\ee
$\Lambda$ is the dynamical scale of the theory.

The v.e.v. of the scalar componet $\phi$ of $A$, determines the moduli space 
of the Coulomb phase of the theory, which is in turn parametrized by the 
invariants of the gauge group.
\be
<\phi>=\sum a_{i} H_{i}
\ee
with $H_{i}$ the Cartan generators.

In a generic point of the moduli space, the gauge group will be broken to
$U(1)^{r}$, where $r$ is the rank of the group and all other gauge fields 
will become massive. However, on a singular set, there will be an 
enlargement of the symmetry group and correspondingly an extra set of gauge 
fields become massless, classically.

There are other states in the theory ,the so called BPS states, with masses
\be
M^{2}=2|Z|^2 =2|\sum_{i} q_{i} a_{i} + g_{i} a^{D}_{i}|^{2} 
\ee
where $a^{D}_{i}={\partial {\cal F} \over \partial a_{i}}$, which become 
massless
on certain other singularities of the quantum moduli space. Here $q_{i}$ 
and $g_{i}$ are the electric and magnetic charges of the dyons.

The important discovery in the N=2 gauge theories has been the realization
that the prepotentials can be described with the aid of a family of complex
curves, with the identification of the v.e.v., $a_{i}$ and their dual 
$a^{D}_{i}$, with the periods of the curve ,
\be
a_{i} = \oint _{\alpha _{i}} \lambda \;\;\;\; \mbox{and} \;\;\;\;
a^{D}_{i} = \oint _{\beta_{i}} \lambda  . \label{integrals}
\ee
where $\alpha_{i}$ and $\beta_{i}$ are the homology cycles of the 
corresponding Riemann surface.

The curves and the Riemann surfaces have been found for the $A_{n}, B_{n}$ and
$D_{n}$ simple Lie groups [2-5]. The importance of the exceptinal groups for
the phenomenology of GUTS and string theory, and also for the possibility of 
a unified description of all N=2 gauge theories, requires knowledge of 
the curve for these groups. In the case of N=1 supersymmetric Yang-Mills 
theories certain results for the exceptinal groups 
have been found, in particular for
$G_{2}$ group \cite{GRE}. In this letter we will present the curve 
for N=2 supersymmetric $G_{2}$ gauge theory and hope to return to the other 
exceptinal groups in a forthcoming work.

The group $G_{2}$ is of rank two and dimension 14. We choose the two simple 
roots as in figure 1.

\begin{center}
\unitlength=1mm
\special{em:linewidth 0.4pt}
\linethickness{0.4pt}
\begin{picture}(38.00,13.00)
\put(10.00,10.00){\circle*{5.20}}
\put(35.00,10.00){\circle{6.00}}
\put(12.00,10.00){\line(2,0){20}}
\put(11.00,12.00){\line(2,0){22}}
\put(11.00,8.00){\line(2,0){22}}
\put(10.00,3.00){\makebox(0,0)[cc]{$1$}}
\put(35.00,3.00){\makebox(0,0)[cc]{$2$}}
\end{picture}
\end{center}
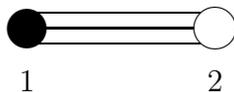
\centerline{ Figure 1 :{\it $G_{2}$ Dynkin diagram }}
\vspace*{4mm}
The Weyl group of $G_{2}$ is the permutation group of order 3 with inversion,
generated by $r_{1}$ and $r_{2}$,
\bea
r_{1} :\,\,  (a_{1},a_{2})  & \rightarrow & (3a_{2}-a_{1},a_{2}) \\
r_{2} :\,\,  (a_{1},a_{2})  & \rightarrow & (a_{1},a_{1}-a_{2})
\eea
The Casimirs of $G_{2}$, written in term of $a_{1}$ and $a_{2}$ are:
\bea
u & = &  a_{2}^{2}+(a_{1}-a_{2})^{2}+(a_{1}-2a_{2})^{2} \cr
v & = &  a_{2}^{2}(a_{1}-a_{2})^{2}(a_{1}-2a_{2})^{2}
\eea

We will take the hyper-elliptic curve for $G_{2}$ to be of the form
\be
y^{2}=(W(x))^{2}-\Lambda^{2h} x^{k}.
\ee
The power of $\Lambda$, twice the dual Coxeter number $h$, which is equal 
to 4 for $G_{2}$, is determined by the $U(1)_{R}$ anomaly \cite{SI}, and $k$
is to be determined by the order of $W(x)$, the classical curve for $G_{2}$.

$W(x)$ is to reflect the singularity structure of the classical theory, which
is determined by the Weyl group. We will therefore construct $W$ in
such a way as to have its discriminant vanish at the Weyl chambers' walls.
Proceeding in this manner, and requiring $W$ to be Weyl invariant, we find:
\be
W=(x^{2}-a_{2}^{2})(x^{2}-(a_{1}-a_{2})^{2})(x^{2}-(a_{1}-2a_{2})^{2}).
\ee
Therefore we get the curve (10) with $k=4$. The quantum discriminant is:
\be
\Delta_\Lambda=\prod_{i<j}(e_i^\pm-e_j^\pm)^2=\Lambda^{72} \Delta^{+}
\Delta^{-} 
\ee
where
\be
\Delta^{\pm}=64v[27({u^3 \over 108}-v\pm {u\Lambda^4 \over 3})^{2}-
4({u^{2} \over 12} \mp \Lambda^{4})^{3}]^{2}
\ee

The branch cuts of the Riemann surface corresponding to the curve (10) 
are depicted in figure 2.

\begin{center}
\unitlength=1mm
\special{em:linewidth 0.4pt}
\linethickness{0.4pt}
\begin{picture}(72.00,59.00)
\put(20.00,35.00){\circle{14.00}}
\put(65.00,35.00){\circle{14.00}}
\put(30.00,13.00){\circle{14.00}}
\put(55.00,13.00){\circle{14.00}}
\put(30.00,52.00){\circle{14.00}}
\put(55.00,52.00){\circle{14.00}}
\put(26.00,55.00){\line(1,-1){7}}
\put(52.00,49.00){\line(1,1){7}}
\put(60.00,35.00){\line(1,0){10}}
\put(15.00,35.00){\line(1,0){10}}
\put(27.00,10.00){\line(1,1){7}}
\put(52.00,17.00){\line(1,-1){7}}
\put(30.00,63.00){\makebox(0,0)[cc]{$\gamma_3$}}
\put(55.00,63.00){\makebox(0,0)[cc]{$\gamma_2$}}
\put(74.00,43.00){\makebox(0,0)[cc]{$\gamma_1$}}
\put(11.00,43.00){\makebox(0,0)[cc]{$\gamma'_1$}}
\put(30.00,2.00){\makebox(0,0)[cc]{$\gamma'_2$}}
\put(55.00,2.00){\makebox(0,0)[cc]{$\gamma'_3$}}
\end{picture}
\end{center}
\vspace*{3mm}
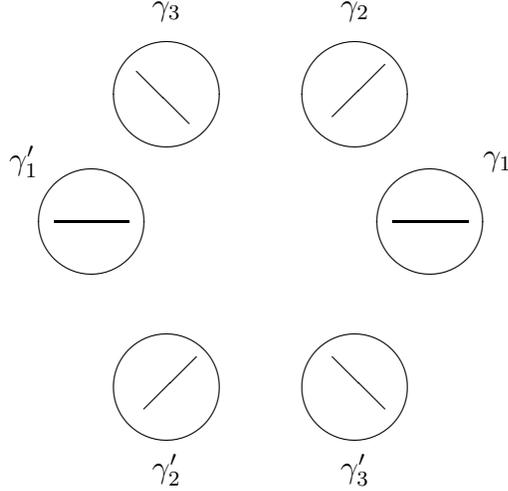
\centerline{Figure 2 :{\it Branch cuts for the curve eq. (10) and the 
$\alpha$ cycles}}
\vspace*{4mm}
Where we have also indicated the corresponding cycles with $\gamma_{i}$ and
$\gamma'_{i}$. The two sets $\gamma_{i}$ and $\gamma'_{i}$ are related by
the parity $x \rightarrow -x$.

To relate  $\gamma_{i}$ to $\alpha_{i}$ of the equation (6), we observe
that in the limit $\Lambda \rightarrow  0$ the vanishing cycles 
should reproduce the classical singularities of the theory \cite{LER}, 
leading to:
\bea
\gamma_{1}&=& \alpha_{2} \cr
\gamma_{2}&=& 2\alpha_{2}-\alpha_{1} \cr
\gamma_{3}&=&  \alpha_{2}-\alpha_{1}
\eea

The intersection requirements of the cycles $\beta_{i}$ with $\alpha_{i}$,
determine the $\beta_{i}$ to be
\bea
\gamma^{D}_{1}&=& \beta_{2}+2\beta_{1} \cr
\gamma^{D}_{2}&=& -\beta_{1}
\eea
where $\gamma^{D}_{i}$ are the conjugate cycles to $\gamma_{i}$
 
It is then straightforward to compute the monodromies for the singularity
 at $\Lambda \rightarrow 0$  and obtain:\footnote{In accordance with the
 notation of Ref[3]}
\be\ba{lll}
B_{1}=\pmatrix{1 &0 &0& 0 \cr 0 & 1 & 0 &-1 \cr 0 & 0 & 1& 0 \cr 0 & 0 & 0& 1}& 
B_{2}=\pmatrix{1 &0 &-1& 2 \cr 0 & 1 & 2 &-4 \cr 0 & 0 & 1& 0 \cr 0 & 0 & 0& 1}&   
B_{3}=\pmatrix{1 &0 &-1& 1 \cr 0 &1 & 1 &-1 \cr 0 & 0 & 1& 0 \cr 0 & 0 & 0& 1}.

\ea\ee
where we have written the matrices in the $(\alpha_{i},\beta_{i})$ basis.
By multiplying these matrices, we will obtain the quantum shift matrix
$T^{-1}$, where
\be
T=\pmatrix{1 &0 &2&-3 \cr 0 & 1 & -3 & 6 \cr 0 & 0 & 1& 0 \cr 0 & 0 & 0& 1}.
\ee
in agreement with the shift obtained from (3) under $\Lambda^{8} \rightarrow
e^{2\pi it} \Lambda^{8}$, $ t \, \epsilon \,[0,1]$.
The semi-classical monodromies obtained from the one-loop corrected 
prepotential of equation (3) are
\be\ba{ll}
M^{(r_{1})}=\pmatrix{-1 &0 &-4& 5 \cr 3 &1 & 7 &-9 \cr 0 & 0 & -1&3 \cr 
0 & 0 & 0& 1}& 
M^{(r_{2})}=\pmatrix{1 &1 &-1& 1 \cr 0 & -1 & 3 &-4 \cr 0 & 0 & 1& 0 \cr 
0 & 0 & 1&-1}\\
M^{(r_{3})}=\pmatrix{2 &1 &-1&4 \cr -3 &-2&2 &-9 \cr 
0 & 0 & 2& -3 \cr 0 & 0 & 1& -2},&
M^{(r_{4})}=\pmatrix{-2&-1&-1&-2 \cr 3 & 2 & 4& -1 \cr 
0 & 0 & -2& 3 \cr 0 & 0 & -1& 2}\\ 
M^{(r_{5})}=\pmatrix{-1 &-1&-1&3 \cr 0 & 1 & -3 &0 \cr 
0 & 0 & -1& 0 \cr 0 & 0 &-1& 1}&
M^{(r_{6})}=\pmatrix{1 &0 &0& 3 \cr -3 & -1 & -3 &-1 \cr 
0 & 0 & 1& -3 \cr 0 & 0 & 0&-1}.
\ea\ee
where $r_3=r_2r_1r_2^{-1}, r_4=r_1r_2r_1^{-1}, r_5=r_3r_1r_3^{-1}
, r_6=r_4r_2r_4^{-1}$.

To calculate the exact monodromies, we look at the vanishing cycles of the
Riemann surface, as the branch points coalesce with $u$ varying, and 
then use Picard-Lefshetz formula. The result is
\be\ba{ll}
M_{1}=M_{(1,0,1,-1)}, & \,\, M_{2}=M_{(1,0,-1,2)}\cr 
M_{3}=M_{(0,1,-1,1)},  & \,\,M_{4}=M_{(0,1,0,-1)}  \cr
M_{5}=M_{(1,1,0,1)}, & \,\, M_{6}=M_{(1,1,1,-2)}\cr
M_{7}=M_{(3,1,1,2)}, & \,\, M_{8}=M_{(3,1,0,3)}\cr 
M_{9}=M_{(2,1,1,-3)},  & \,\,M_{10}=M_{(2,1,0,-3)}  \cr
M_{11}=M_{(3,2,3,-2)}, & \,\, M_{12}=M_{(3,2,3,-3)}.

\ea\ee
In the above equation we have used the notation\cite{KE}
\be
M_{(\m,\qq)}=\pmatrix{\bone-\qq \otimes \m & - \qq \otimes \qq \cr
\m \otimes \m & \bone+\m \otimes \qq }.
\ee

As a check of the curve (10), we can reproduce the semi-classical monodromies 
by multiplying pairs of the above exact monodromies, i.e.
\be\ba{lll}
M_2 M_1 =M^{(r_{1})}, & M_4 M_3=M^{(r_{2})}, & M_6 M_5=M^{(r_{3})} \\
M_8 M_7 =M^{(r_{4})}, & M_{10} M_9=M^{(r_{5})}, & M_{12} M_{11}=M^{(r_{6})}
\ea\ee
It is interesting to note that if we consider the map $ S $
\be
\pmatrix{\bde^{D} \cr \bde}= S\,\, \pmatrix{\bal^{D} \cr \bal}.
\ee
with
\be
S=\pmatrix{1 &1 &0& 0 \cr -2 & -1 & 0 & 0 \cr 0 & 0 & -1& 2 \cr 0 & 0 & -1& 1}.
\ee
then the subset $(M_1, M_2, M_5, M_6, M_9, M_{10})$ of eq. (19) monodromies,
transformed as $M'_i=S^{-1}\,M_i\,S $ turn out to be 
those of N=2 gauge theory with gauge group $SU(3)$. 
The subset of the dyons (19) are correspondingly mapped into the dyons of 
the $SU(3)$ theory generated by the short roots \cite{KE,LA}. 

There is an interesting connection between the theories with $G_{2}$ and $SU(3)$ gauge
groups which can be seen by the following change of variable in the
curve of $G_{2}$ (10),
\be\ba{ll}
x^{2}=x'+{u \over 3},  & \,\,\, \Lambda'=\Lambda^{2}.
\ea\ee
giving
\be
y'^2 = (x'^{3}-u' x'-v')^{2} - \Lambda'^{4} (x'+{u \over 3})^{2}.
\ee
We recognize this curve to be that of the N=2 supersymmetric $SU(3)$ gauge
theory with two massive hypermultiplets in the fundamental 
representation \cite{HAN}.

Our $G_{2}$ curve may shed further light on the current attempts at a  
better understing of the exact results for $N=2$ supersymmetric Yang-Mills
theories. For example there have been attempts to cast these results in terms 
of the integrable systems \cite{MOR}. In a related approach, it has been shown
that,\cite{EG}, the prepotential of the
N=2 gauge theory with the gauge groups $A_{n},B_{n}$ and $D_{n}$ satisfy the
solitonic equation
\be
\sum_{i}^{r} a_{i} {\partial {\cal F} \over \partial a_{i}}-2{\cal F}=
8 \pi i b_{1} u
\ee
where $b_{1}$ is the coefficient of the one-loop $\beta$ funtion. 
To verify whether this same equation is satisfied by our curve (10), 
we note that equation (26) was derived from \cite{EG}
\be
\sum_{i}^{r} a_{i} {\partial {\cal F} \over \partial a_{i}}-2{\cal F}=
- T_{1} {\partial {\cal F} \over \partial T_{1}}
\ee
with the assumption that the curve is invariant under $x \rightarrow -x$.
Here $T_{1}$ and ${\partial {\cal F} \over \partial T_{1}}$ are the 
coefficient
in an asymptotic expansion of the one form $\lambda$ in powers of 
$z={1 \over x}$
as $x \rightarrow \infty $
\be
\lambda=(-\sum_{n \geq 1} n T_{n}z^{-n-1}+T_{0}z^{-1}
-{1 \over 2 \pi i}\sum_{n \geq 1} {\partial {\cal F} \over \partial T_{n}} 
z^{n-1})dz.
\ee
In the case of $G_{2}$, $\lambda$ is
\be
\lambda=(-4 x^{6} +2u x^{4}-2v) {dx \over 2\pi i  y}
\ee
Reading the coefficient $T_{1}$ and ${\partial {\cal F} \over \partial T_{1}}$ 
>from the expansion of equation (29), we find
\be\ba{ll}
T_{1}={4 \over 2\pi i}, \,\,\, & {\partial {\cal F} \over \partial T_{1}}= 2u
\ea\ee
Thus verifying the solitonic equation (26).

After completion of this work similar results appeared in the
work by U. H. Danielsson and B. Sundborg.\cite{SU}

\vspace*{8mm}

{\large ACKNOWLEDGEMENTS}

M. A. and F. A. would like to thank Hessam Arfaei and Cesar Gomez for 
helpful discussions and Shahrokh Parvizi and Masoud Ghezelbash for 
collabration in the early stages of this work. We greatly appreciate
referee's comments and suggestions. This work was supported 
in part by the Department of Energy under the contract number 
DOE-FG02-84ER40153.

\end{document}